\documentclass[aps,twocolumn,groupedaddress]{revtex4}
\usepackage{soul}
\usepackage{physics}

\usepackage{graphicx}

\def\presuper#1#2%
{\mathop{}%
	\mathopen{\vphantom{#2}}^{#1}%
	\kern-\scriptspace%
#2}

\usepackage[utf8]{inputenc} 

\usepackage{paralist}
\usepackage{wrapfig}
\usepackage{gensymb}
\usepackage{tabularx}
\usepackage{capt-of}
\usepackage{color}
\usepackage{subfigure}
\usepackage{multirow}
\usepackage{array}
\graphicspath{{images/}}
\usepackage{amsmath,amssymb,amsfonts,textcomp}
\usepackage{mathtools}

\usepackage{color}
\usepackage{subfigure}
\usepackage{multirow}
\usepackage{array}
\usepackage{outlines}    

\usepackage{hyperref}   
\setcitestyle{super}

\newcolumntype{R}{>{\raggedleft\arraybackslash}X}

\newcounter{ctComment}
\begin{document}


\title{Coupled electron pair-type approximations for tensor product state wavefunctions} 


\author{Vibin Abraham}
\email{avibin@umich.edu}
\affiliation{Department of Chemistry, University of Michigan, Ann Arbor, MI 48109, USA}
\author{Nicholas J. Mayhall}
\email{nmayhall@vt.edu}
\affiliation{Department of Chemistry, Virginia Tech,
Blacksburg, VA 24060, USA}

\begin{abstract}

Size extensivity, defined as the correct scaling of energy with system size, 
    is a desirable property for any many-body method.
Traditional CI methods are not size extensive hence the error increases as the system gets larger.
Coupled electron pair approximation (CEPA) methods can be constructed as simple extensions of truncated configuration interaction (CI) 
     that ensures size extensivity.
One of the major issues with the CEPA and its variants is that singularities arise in the amplitude equations 
    when the system starts to be strongly correlated.
In this work, we extend the traditional Slater determinant-based coupled electron pair approaches like CEPA-0,
    averaged coupled-pair functional (ACPF) and average quadratic coupled-cluster (AQCC) to
    a new formulation based on tensor product states (TPS).
We show that a TPS basis can often be chosen such that it removes the singularities that commonly destroy the accuracy of CEPA-based methods. 
A suitable TPS representation can be formed by partitioning the system into separate disjoint clusters and 
    forming the final wavefunction as the tensor product of the many body states of these clusters.
We demonstrate the application of these  methods on simple bond breaking systems such as CH$_4$ and F$_2$ 
    where determinant based CEPA methods fail.
We further apply the TPS-CEPA approach to stillbene isomerization and few planar $\pi$-conjugated systems.
Overall the results show that the TPS-CEPA method can remove the singularities and provide improved numerical results compared to common electronic structure methods.
\end{abstract}

\maketitle

\section{Introduction}
Accurate estimation of the total energy of the system is of prime importance to quantum chemistry.
When high accuracy and systematic improvability is necessary, wavefunction based methods like coupled cluster (CC), perturbation theory (PT), and configuration interaction (CI) are typically the preferred approaches.
However, starting from a single Slater determinant reference leads to challenges when no single determinant can qualitatively describe the electronic structure. 
Such strongly correlated states show up in a wide range of important situations, e.g., transition metal complexes, excited states, transition states, etc. 
Despite the importance of such systems, there is currently no polynomially scaling technique which is effective at treating strong correlation for a general system. 

The typical approach to modeling strongly correlated systems is to use a proper ``multi-reference'' method. 
These are usually constructed around the concept of an orbital active-space,\cite{roos1980complete,veryazov2011select,ruedenberg1982atoms} 
one which is chosen to capture the strong correlation effects. 
However, since strong correlation is a property of the many-electron basis, it is not always possible to find a small number of single-electron functions that qualitatively characterize the strong correlation. 
While the active-space concept is often useful in providing a qualitatively correct description of strong correlation, one usually needs to go beyond this model for quantitatively correct results.

Truncated CI is an intuitive and simple framework for the remaining correlation energy on top of a multiconfigurational reference. 
Multireference CI,\cite{buenker1978applicability,shavitt1998history} is improvable by including higher excited configurations systematically.
Though conceptually simple and physically intuitive, truncated CI methods lack the property of size extensivity, which is defined as the correct scaling of energy with system size.\cite{Bartlett1981,Nooijen2005,shavitt_bartlett_2009}
This becomes important as we go to larger and larger systems since the energy error due to lack of size extensivity increases.
A simple approach to correct for the size extensivity is by \textit{a posteriori} corrections computed using the converged CI coefficients.\cite{langhoff1974configuration,bruna1981non,pople1977variational,meissner1988size}
While useful and simple, these corrections are rather ad hoc, and still don't produce exactly size-extensive results. 
Unlike CI, coupled cluster (CC) is size extensive through the use of an exponential ansatz.
The cost of this is that the coupled cluster equations are non-linear, and since rooted in a diagrammatic approach, 
    extension to multi-reference is neither straight forward nor numerically stable.\cite{Evangelista2018}
Predating CC, coupled electron pair approximation (CEPA) methods are another family of methods that, like CC, can be formulated to be strictly size extensive. 
However, similar to CI, CEPA methods constitute a linear parameterization.\cite{meyer1973pno,AHLRICHS1979}
This linear nature makes CEPA methods much more extendable to multireference formulations which are both simpler and more numerically stable than the MR-CC approaches.\cite{fink_multi-configuration_1993,chattopadhyay2004state,SZALAY2008,malrieu1994multireference}
The computation of analytic gradients is  also often straight forward for these linear approaches.\cite{lischka2001high,Taube2009,Bozkaya2013,lischka_analytic_2002,szalay2000excitation}

There are many distinct CEPA formulations.
In the most common approach (CEPA-0), the exclusion principle violation (EPV) terms are neglected and hence provides size extensive results.\cite{AHLRICHS1979}
CEPA-0\cite{AHLRICHS1979,koch_comparison_1981,Kollmar2010} has connections to other methods like DMBPT$\infty$ method by Bartlett,\cite{BARTLETT1977}
    linearized coupled cluster (LCC) methods,\cite{Bartlett1981,Taube2009}
    and the parametric variational 2RDM method.\cite{deprinceParametricApproachVariational2007,Vu2020}
Because CEPA-0 can be written as the variational minimization of an approximate energy expression, this method often results in the overestimation of the correlation effects arising from the neglect of the EPV terms.

In the averaged coupled-pair functional (ACPF) approach, the EPV terms are considered by averaging the correlation energy over the number of electron pairs.\cite{gdanitz1988averaged,Venkat2004}
Even though the equations are not strictly size-extensive this approximation works very well for binding energies and transition moments, etc.\cite{rezabal2011quantum,szalay2000excitation}
In particular, the single reference version becomes exactly size-consistent for the separation of identical subsystem.\cite{gdanitz1988averaged}
Another approach is the average quadratic coupled-cluster (AQCC) which can be understood as a damped ACPF approach.\cite{meissner1988size,SZALAY1993,szalay1995approximately,szalay2000excitation,FUSTIMOLNAR1996}

Other approaches exist, like CEPA-n (n=1, 2\cite{meyer1971ionization,meyer1973pno}, 3\cite{kelly1963correlation}) where the EPV terms are approximated using the orbital pair energies. 
Because these methods depend on the orbital basis, they lack orbital rotational invariance.\cite{szalay2005configuration,WENNMOHS2008,ruttink2005multireference}
The self-consistent size-consistent CI ((SC)$^2$CI) includes all exclusion principle violating (EPV) terms,\cite{daudey1993size,junquera2006size}
but is more expensive than the other CEPA methods mentioned and is also not invariant under orbital rotations.

In addition to approaches that incorporate additional EPV terms, other strategies to improve CEPA have been explored.
Nooijen and coworkers developed a size extensive way to include approximate triples to CEPA. \cite{NOOIJEN2006}
Bozkaya and workers have demonstrated that orbital optimization can significantly improve CEPA(0), providing results more accurate than CCSD for hydrogen transfer reactions. \cite{Bozkaya2013}
Sharma and Alavi designed a linearized coupled-cluster approach based on matrix product states.\cite{sharmaMultireferenceLinearizedCoupled2015a}
The MR-AQCC approach has been demonstrated to produce accurate excitation energies\cite{szalay2000excitation} and also has provided one of the most qualitatively accurate potential energy curve for the dissociation of chromium dimer, a challenging system frequently used to benchmark strong correlation methods.\cite{muller2009large}


Although CEPA methods can often be surprisingly accurate, 
they are also surprisingly sensitive to strong correlation.
The main challenge in  single reference CEPA arises from near degeneracies.
For example, a simple single bond breaking only requires doubly excited configurations to obtain a qualitatively correct behavior.
Despite containing all the necessary configurations,
CEPA-0, ACPF and AQCC  fail to produce qualitatively correct results.\cite{MALRIEU2010,Taube2009}
Coupled cluster meanwhile works well when sufficient determinants are present.
Hence one needs to be cautious when using CEPA in practice.
This sensitivity can be reduced by introducing a regularization\cite{Taube2009} or by using other CEPA variants that depend on orbitals like CEPA-3 or (SC)$^2$CI.\cite{MALRIEU2010}

\subsection{Course-grained quantum chemistry}
Although molecules cannot be simply thought of as merely collections of atoms in close proximity, many atomic, or local properties persist in a molecule and transfer between molecules. 
For example, an alcohol group behaves similarly regardless of whether it's attached to pentane or hexane. 
This transferability of local character between different systems implies a certain degree of cluster-based low-entanglement. 
Our group has recently begun exploring the ability to use tensor product state representations to encode and exploit transferable local structure.\cite{abrahamClusterManybodyExpansion2021, abrahamSelectedConfigurationInteraction2020c, mayhallUsingHigherOrderSingular2017a}    
We have been developing new methods by forming the many-body wavefunction of the full system in the basis of tensor products of many-body states of disjoint groups of orbitals, or clusters.
Recently, we have demonstrated that a selected CI procedure performed in a TPS basis requires fewer parameters than when performed in the traditional Slater determinant basis, 
and that extremely high accuracy can be obtained for systems that exhibit a clusterable structure.\cite{ abrahamSelectedConfigurationInteraction2020c}   
Even the many body expansion based approach has better convergence for TPS as compared to Slater determinants.\cite{abrahamClusterManybodyExpansion2021}
Although the methods realized in a TPS basis can be much more compact and more computationally efficient than when using a determinant basis, the TPS approach is not a black-box method, and the user needs to carefully choose the orbital clustering for each application.
This is analogous to (but more involved than) the choice of active space in CASSCF or orbital ordering in DMRG. 

For a given system, once the disjoint orbital clusters are defined,
    we can perform a mean field optimization, varying both the molecular orbitals defining the clusters as well as the many-body cluster ground states to obtain the variational, lowest energy single TPS.
This is simply the cluster mean-field (cMF) approach developed by Jim\'enez-Hoyos and Scuseria,\cite{Jimenez-Hoyos2015}
or the vLASSCF method from Hermes and coworkers.\cite{Hermes2020} 
For clusterable systems, the cMF is a good starting point because all intra-cluster interactions are included explicitly, while inter-cluster interactions are included at a mean-field level (more specifically, through interactions with a one-particle reduced density matrix ). 
By defining a clustering in which all strong correlations are contained within a cluster, the cMF reference is an accurate reference state for further improvement.
Our recently developed tensor product selected CI (TPSCI) method, starts with cMF and includes the inter-cluster interactions by diagonalizing the Hamiltonian in a basis of excited tensor product states, a basis which is constructed by an iterative selected CI-like procedure.\cite{abrahamSelectedConfigurationInteraction2020c}

While TPSCI was found to quickly converge to FCI, it is an adaptive method, and as such, inherits the difficulties of other adaptive methods. 
When computing PES curves, discontinuities arise to inconsistent truncation.
In contrast, non-adaptive methods can be useful far from the FCI limit due to the fact that systematic errors largely cancel, yielding accurate relative energies even when the absolute energies are far from converged. 
Adaptive methods don't typically benefit from error cancellation, and so much higher absolute accuracy is needed. 

Non-adaptive methods in a TPS basis have been explored by Li and coworkers with the block correlated coupled cluster method (BCCC),\cite{liBlockcorrelatedCoupledCluster2004} excitonic coupled cluster (XCC)\cite{Liu2019,Liu2019b} by Dutoi \textit{et al} and the recently proposed cluster-CCSD (cCCSD) by Scuseria.\cite{papastathopoulos2021cluster}
In each of these approaches, inter-cluster correlations are included with either perturbation theory or coupled cluster theory. 
Whereas BCCC was the first realization of this approach, the reference TPS state was not self-consistently optimized until Scuseria and coworkers applied PT2 and CC on top of the cMF reference.\cite{Jimenez-Hoyos2015,papastathopoulos2021cluster}
We refer to these as post-cMF methods, 
analogous to post-SCF terminology.
A few coarse grain approaches have been introduced using the density matrix renormalization group (DMRG) framework as well: the active space decomposition ASD-DMRG, \cite{parkerCommunicationActiveSpace2014}
 comb-DMRG,\cite{li2021expressibility} and the multi-site matrix product states method.\cite{larsson2022matrix}

In the present work, we explore new \textit{ab initio} post-cMF methods in the framework of CEPA
and demonstrate advantages over Slater determinant based approaches. 



\section{Theory}\label{sec:theory}
Traditional wavefunction-based methods start from a mean field determinant and expand the wavefunction as excitations from this reference.
In the TPS framework one can do the same but from a mean field tensor product state wavefunction.
In this approach, the active space of the system is partitioned into disjoint orbital sets.
We can solve the many body problem in each of these smaller clusters and represent the wavefunction as a tensor product of these cluster states.
We can further optimize the orbitals and the cluster state many-body states variationally for a single TPS state.
The zeroth order operator, or the TPS Fock-like operator, is given as:
\begin{equation}
\hat{H}^0 =\sum_{I} \hat{H}_I  + \sum_{I,J} \sum_{pr\in I}\hat{p}^{\dagger}\hat{r}  \sum_{qs \in J}    \mel{pq}{}{rs} \rho^{J}_{qs}  
\end{equation} 
where upper-case roman letters, $I$/$J$, enumerate clusters, lower-case letters refer to orbitals, and $\rho^{J}_{qs} = \mel{0_J}{\hat{q}^{\dagger}\hat{s}}{0_J}$ is a one-particle density matrix for the lowest energy cluster state in $J$. 
The reference cMF state can be represented as 
\begin{equation}
    \ket{\Phi_0} =  \ket{0_{I}}\ket{0_{J}},..\ket{0_{N}} = \ket{0_{I},0_{J},..0_{N}},
\end{equation}
where we use the notation $\ket{0_K}$ as the ground state of cluster $K$.

While including all intra-cluster interactions, cMF reference still lacks correlation between clusters, and this correlation can 
    be captured using higher excited TPS configurations. 
Jim\'enez-Hoyos and co-workers used a simple perturbation theory to capture this correlation.\cite{Jimenez-Hoyos2015}
Block correlated coupled cluster (BCCC)\cite{liBlockcorrelatedCoupledCluster2004,Wang2020}, excitonic coupled cluster (XCC)\cite{Liu2019,Liu2019b} and the recently proposed cluster-CCSD (cCCSD)\cite{papastathopoulos2021cluster} use the coupled cluster ansatz in a TPS framework.
The BCCC uses a generalized valence bond (GVB) reference while the XCC approach uses an approximate basis formed using a local Hamiltonian as the reference.\cite{Wang2020}
The cCCSD method uses the cMF reference but has not yet been extended to molecules.
Similar to determinants, these coupled cluster based approaches in the TPS framework would also be much harder to extend to a multi-reference framework.

In order to develop the TPS-CEPA methods, we start by defining the simpler TPS-CI wavefunction, 
which is a simple linear parameterization of the (potentially) exact wavefunction.
By taking the cMF wavefunction as our reference state, $\ket{\Phi_0}$,
the CI wavefunction using intermediate normalization can be written as a linear combination of excited TPS configurations:
\begin{equation}
    \ket{\Psi} = \ket{\Phi_0} + \sum_{I} c_I \ket{\Phi_I},
\end{equation}
where $\ket{\Phi_I}$ is an excited TPS configuration and $c_I$ are the corresponding coefficients. 

In contrast to the Slater determinant basis for which the first order interaction space (FOIS) is defined by single and double excitations,
in the TPS basis, the FOIS contains  up to quadruple excitations (though not all quadruple excitations). 
For example, when the indices of a two electron term, $\hat{p}^\dagger\hat{q}^\dagger\hat{r}\hat{s}$, occur in four distinct clusters, those four clusters are simultaneously excited into electron attached/detached states. 
Methods like XCC, BCCC and cCCSD define their target spaces based on cluster excitation rank (i.e., the number of clusters simultaneously excited),  usually truncating to doubles. 
In order to fully span the FOIS using excitations in this framework, one would require quadruple excitations, leading to a much larger number of configurations needed.
Instead of defining the FOIS by tedious labelling of cluster Fock space configurations,  one can directly form the FOIS for the TPS framework by applying the Hamiltonian to the reference cMF wavefunction.
This leads to a more compact space which will have selective configurations from the doubles, triples and quadruple excitation of the TPS framework. 
Hence, we generate our target space, $\ket{\Phi_I}$, by generating the FOIS of the Hamiltonian in the TPS basis.
\begin{align}
    \hat{H}\ket{\Phi_0} =& \sum_{I} \ket{\Phi_I}\mel{\Phi_I}{\hat{H}}{\Phi_0} \\
    =& \sum_{I} \ket{\Phi_I}h_I + E_0\ket{\Phi_0} = \ket{Q}
\end{align}

Because most of the excitations will have a zero matrix element, $\mel{\Phi_I}{\hat{H}}{\Phi_0}$, 
this is much smaller than the full excitation space up to quadruples in the TPS basis. 
The exclusion of excitations outside of the FOIS is important for reasons other than just reducing computational cost. 
CEPA based methods are generally only defined for the FOIS, and the inclusion of excitations not in the first order interacting space can destroy performance and size extensivity.

Given the exact ground state, $\ket{\Psi}$, the exact post-cMF correlation energy  can be computed as, 
\begin{align}
    E_c =& \mel{\Phi_0}{\hat{H}-E_0}{\Psi}\nonumber\\
    =& \sum_{I}h_Ic_I= 
    \mathbf{h}^\dagger\mathbf{c}\label{eq:ecorr}.
\end{align}
The coefficients of each TPS configuration can be calculated by projecting the equation onto excited TPS configurations:
\begin{align}
    \label{eq:CI}
    0 =& \mel{\Phi_I}{\hat{H}-E_0-E_c}{\Psi}\nonumber\\
    =&\sum_J \mel{\Phi_I}{\hat{H}-E_0-E_c}{\Phi_J}c_J\\
    =&\mathbf{A}(E_c)\mathbf{c}\label{eq:linsys}.
\end{align}
In this form, one could obtain the ground state coefficients, $\mathbf{c}$, by solving the linear system in Eq. \ref{eq:linsys} for a fixed, $E_c$, then computing a new $E_c$ with Eq. \ref{eq:ecorr}, then solving the linear system again, iterating until convergence.

Because we only need to know the coefficients in the FOIS to
compute the exact energy (Eq. \ref{eq:ecorr}), we now choose to focus only on developing approximations to determining the FOIS coefficients. 
The presence of the correlation energy in  Equation \ref{eq:CI} makes the truncated CI no longer size extensive.
Modifying this equation such that the $E_c$ is replaced by some alternative expression can restore size-extensivity, for example,
\begin{equation}
    0 = \mel{\Phi_I}{\hat{H}-E_0-\Delta}{\Psi},
\end{equation}
where different choices of $\Delta$ lead to the different methods defined in Table, \ref{tbl:delta}.

Ignoring this term leads to the approximation called CEPA-0 (identical to linearized coupled cluster (LCC)).
The CEPA-0 is exactly size extensive but usually overestimates the correlation energy.
ACPF and AQCC methods treat the EPV effects in an averaged manner.
In ACPF, it is approximated that the correlation energy is equally distributed among every electron pair 
    by setting the shift to be $E_c/n_{pairs}$.
Motivated by the Meissner's \cite{meissner1988size} post CI correction for size-extensivity, the AQCC method was proposed with 
    a improved shift.\cite{SZALAY1993}
Both of these methods provide approximately size extensive results and are an improvement over CEPA-0.
For the TPS based approaches, we use the same shifts as in the determinant case.
Similar to other implementations, we employ these approaches by a shift to the diagonal of CI Hamiltonian.

\begin{figure*}
	\begin{center}
	\includegraphics[width=\linewidth]{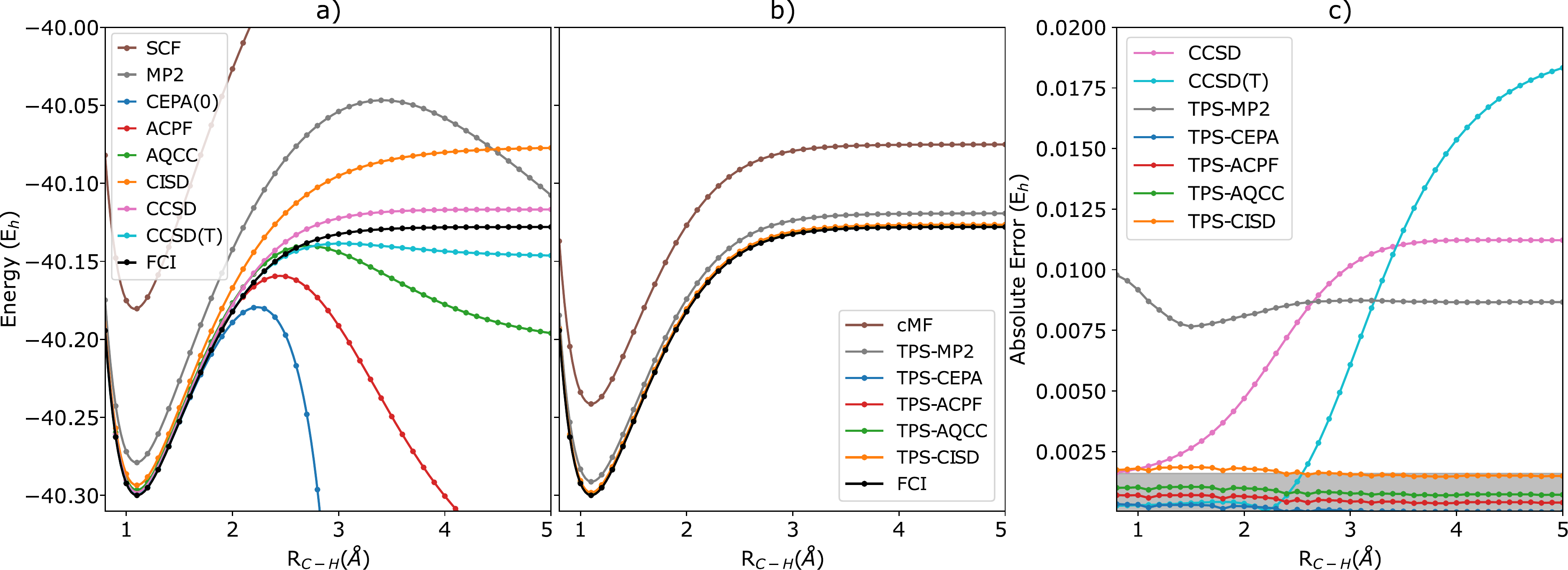}
	\caption{Potential energy surface of the CH$_4$ molecule with respect to one stretched C-H bond using a) determinant basis, b) TPS basis and c) the error with FCI for CCSD, CCSD(T) and the TPS approaches. The grey area corresponds to chemical accuracy.}
	\label{fig:ch4}
	\end{center}
\end{figure*}

\begin{table}
    \caption{The approximate CEPA methods and the shift ($\Delta$) for each. N is the total number of electrons. }
    	\label{tbl:delta}
\begin{tabularx}{0.8\linewidth} {
                                  >{\centering\arraybackslash  \hsize=.4\hsize}X
                                | >{\centering\arraybackslash  \hsize=.6\hsize}X
                                   }
\hline \hline
approximation   & $\Delta$                  		\\ \hline 
CISD$^*$ 		& $E_c$                  		\\ 
CEPA 		&  0 					\\ 
ACPF 		& $\frac{2}{N} E_c$                 	\\ 
AQCC 		& $[1-\frac{(N-3)(N-2)}{N(N-1)}]E_c$	\\ \hline \hline
\end{tabularx}\raggedright\\
$^*$We use CISD to denote CI in the FOIS to keep the analogy to determinant based methods.
\end{table}

As mentioned earlier, there are other flavours of these methods like CEPA(n) where n=1,2,3 and the (SC)$^2$CI methods.
An extensive list can be seen in the review by Wennmohs \textit{et al}.\cite{WENNMOHS2008}
These methods are not invariant under the rotation of the occupied molecular orbitals and hence not studied in this work.
Moreover, correcting the equation with orbital energies is not straight forward since we are working in the TPS basis,
but this could be the focus of future work.


\section{Results}\label{sec:results}
In this section we provide numerical demonstrations of the performance of the TPS-based CEPA methods, drawing comparisons to Slater determinant approaches. 
As an example of traditional strongly correlated test cases where CEPA fails, we study the single bond breaking of CH$_4$ and F$_2$ molecules. 
We then explore larger systems with delocalized electronic structure, including the isomerization of stillbene 
    and a few $\pi$-conjugated systems, specifically a series of polycyclic aromatic hydrocarbons (PAH) and a molecular polymer, polypyrrole.
For the traditional Slater determinant approaches, we used the Psi4 package which contains implementations of numerous CEPA variants.\cite{psi42020}
The TPS-based calculations were performed with our FermiCluster code,\cite{fermicluster} 
using molecular integrals calculated from PySCF.\cite{pyscf2020}
Because this is a prototype implementation (computing matrix elements one-at-a-time)
our numerical tests are limited in size, with algorithmic optimization deferred to a future study.

\subsection{CH$_4$ bond breaking}
Single bond breaking is a classic example of failure for methods like CEPA, ACPF, and AQCC.
For qualitative correctness, a single bond breaking requires up to double excitations and all of the above mentioned approaches (along with CISD and CCSD)
    have these excitations in the determinant basis.
Despite having double excitations,  each of the CEPA approaches fail in these cases.
This does not arise from an absence of important configurations (as would be the case with double or triple bond breaking), but rather due to singularities that arise when there is a near degeneracy.

To demonstrate the failure of the determinant based CEPA, we study the single bond breaking in the CH$_4$ system using a 6-31G basis set.
We fix the angle between the bonds at $109.5\degree$, three of the bonds at 1.086 $\textrm{\AA}$ and vary the length of one C-H bond from 0.8 $\textrm{\AA}$ to 5.0 $\textrm{\AA}$. 
The 1$s$ orbital of the C atom is frozen for all the calculations.
We compute all the determinant based (CISD, CCSD, CCSD(T), and all CEPA variants) using RHF molecular orbitals and compare with the exact FCI results.

As seen in Figure \ref{fig:ch4}(a), most of the determinant based CEPA variants work well in the weakly correlated regime, but ultimately diverge as the bond is stretched.
CEPA is more accurate compared to ACPF and AQCC for the weakly correlated region. 
As the bond is broken, the CEPA method fails first, followed by ACPF and then AQCC. 
CCSD provides a qualitatively good bond breaking curve while CCSD(T) has a small nonphysical bump in the potential energy surface.

In order to define a TPS basis, one must first partition the orbitals into smaller orbital domains, i.e., clusters.
For clustering the valence space, we use intrinsic bonding orbitals.\cite{Knizia2013}
The four C-H bonds are partitioned into separate clusters and the higher excited virtual orbitals are partitioned based on off-diagonal couplings in the core Hamiltonian.
They are partitioned into two clusters of 4 orbitals each.
The clustering scheme is provided in the supporting information. 

In Figure \ref{fig:ch4}(b) we present the PES scan using the TPS based approaches.
The coupled pair methods like TPS-CEPA, TPS-ACPF and TPS-AQCC have negligible error with respect to the FCI curve.
For TPS-MP2, we see that although the absolute error is significantly larger, the relative energies are quite accurate, such that TPS-MP2 provides a qualitatively correct PES curve unlike the determinant based MP2.
We present the error of all the TPS approaches with the FCI in Figure \ref{fig:ch4}(c). 
CCSD and CCSD(T) errors are also plotted for comparison.
The TPS-CEPA and CCSD(T) are very accurate for the weakly correlated regime.
As we go beyond 2.4 $\textrm{\AA}$, the CCSD(T) curve diverges from the FCI curve but the TPS-CEPA errors remain low. 
All three TPS based coupled electron approaches give very accurate results within chemical accuracy (1.6 mH). 

\subsection{F$_2$ bond breaking}
The F$_2$ molecule has one of the weakest covalent bonds, with electron occupations in the anti-bonding orbitals.
The F$_2$ molecule has three lone pairs per atom and has large electronic repulsion, leading to a shallow potential energy surface.\cite{Forslund2003,Bytautas2007,Bytautas2007a}
Accurate wavefunction based approaches\cite{daudey1993size,Ivanov2006,KOWALSKI2001,Bytautas2007,Bytautas2007a,esterhuysen_nature_2004,Evangelista2007,Li2006,Shovan2019,Angeli2006,chatterjee2016minimalistic} are
 necessary to capture the binding energy of F$_2$ molecule due significantly large dynamic correlation.
 Hence it is a good stress-test for new quantum chemistry methods,
 highlighting the ability to capture static, as well as, dynamic correlation.\cite{Ivanov2006,chatterjee2016minimalistic}
 Here, the F$_2$ molecule is studied with varying bond length from 1.2 $\textrm{\AA}$ to 3.0 $\textrm{\AA}$ using the 6-31G basis with the frozen core approximation.

\begin{figure}
	\begin{center}
	\includegraphics[width=\linewidth]{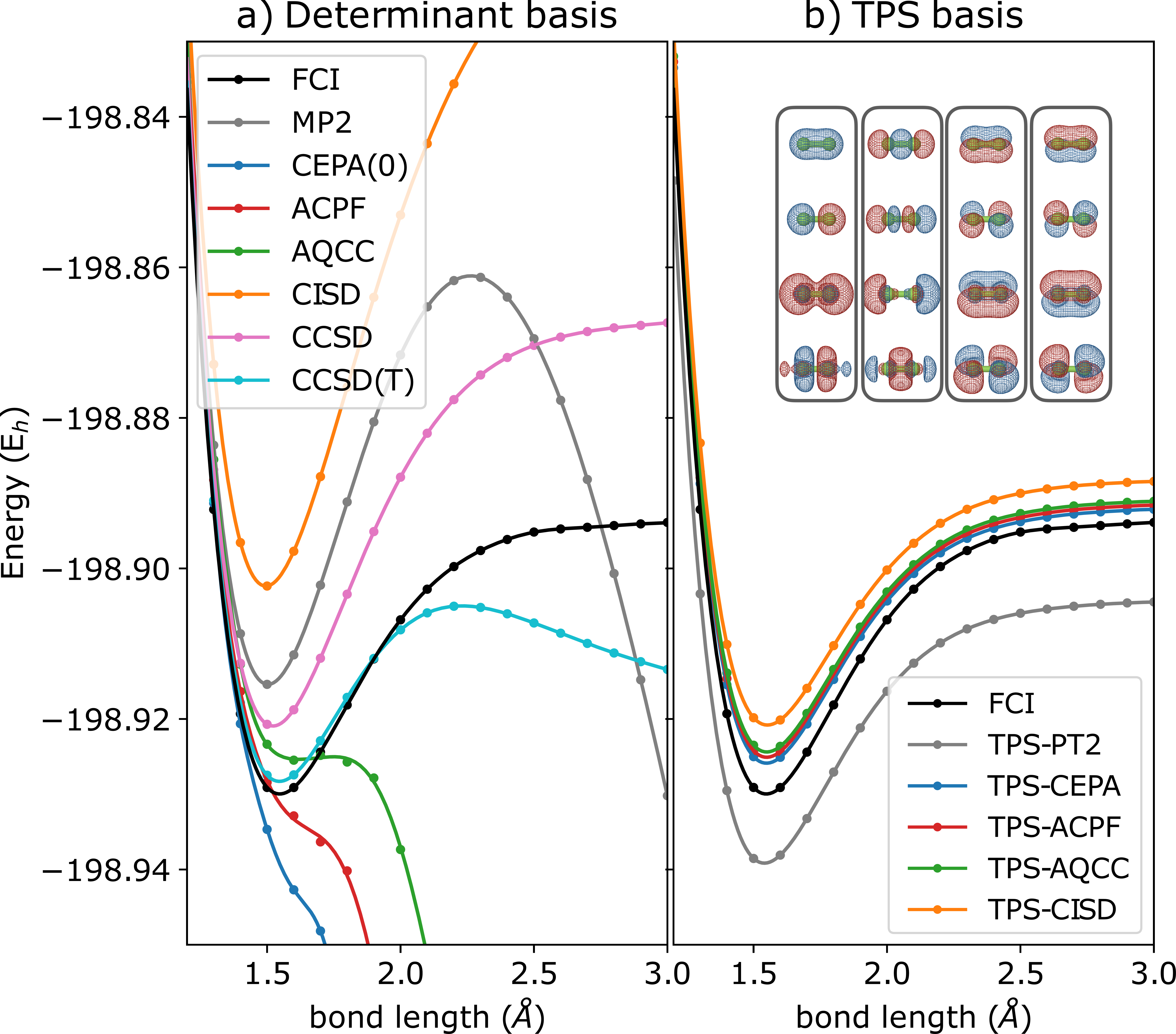}
	\caption{Potential energy surface of the F$_2$ molecule using a) determinant basis and b) TPS basis. The molecular orbital clustering used for the TPS data is also presented.}
	\label{fig:f2}
	\end{center}
\end{figure}

In Figure \ref{fig:f2}(a), we present data for the determinant-based single reference methods
	in breaking the F$_2$ bond.
It can be seen that MP2 as well as all the CEPA variants do not produce meaningful PES. 
Unlike in the case of CH$_4$, the CEPA variants for F$_2$ do not even have well defined minima.
The unphysical bump in the CCSD(T) curve is also more pronounced in F$_2$.
The CCSD approach qualitatively gives a correct PES but is still far from the exact FCI curve.
Large portion of dynamic correlation effects are missing from the CCSD curve as well.

For the TPS basis, we partition the orbitals such that the dynamic correlation part for each valence bonding and anti-bonding orbital is included in the same cluster.
For F$_2$ in the 6-31G basis, we have 16 orbitals and this partitioning leads to 4 clusters with 4 orbitals each:

\begin{align}
\small
&[\sigma_{2s}, \sigma^*_{2s},\sigma_{3s}, \sigma^*_{3s}]
[\sigma_{2p_z}, \sigma^*_{2p_z}, \sigma_{3p_z},  \sigma^*_{3p_z}] \nonumber \\
&[\pi_{2x},  \pi^*_{2x},\pi_{3x}, \pi^*_{3x}]
[\pi_{2y},  \pi^*_{2y},\pi_{3y}, \pi^*_{3y}] \nonumber
\end{align}

We present the TPS based methods for this system in Figure \ref{fig:f2}(b).
The TPS-PT2 method gives qualitatively accurate PES even though it overestimates the correlation energy. 
While the TPS-CISD approach is farthest from the exact curve, the relative energies are quite good. 
However, not being size-extensive, this behavior will worsen for larger systems. 
The size extensive TPS-CEPA, TPS-ACPF and TPS-AQCC methods all have higher accuracy for the F$_2$ PES curves relative to both TPS-MP2 and TPS-CISD.
Because the strong correlation was captured by the cMF reference, these methods do not have any singularity issues, in contrast to the determinant based coupled pair approaches.
Even though the determinant based CCSD(T) is more accurate in the near equilibrium geometry, it starts breaking down at the dissociation limit.
Even though the TPS based methods do not recover the full correlation energy, the relative shape of the PES is reasonable.
We study this using the non-parallelity errors (NPE) as shown in Table \ref{tbl:npe}.
The NPE is defined as the maximum error minus the minimum error relative to the FCI result.
As seen from the table, the NPE errors for all the TPS approaches, are much smaller.
Even the TPS-MP2 has small NPE implying that the PES is of good quality and will lead to meaningful binding energies.
Hence, the TPS based coupled pair methods are much more robust compared to their determinant based alternatives.

\begin{table}[]
\caption{The non-parallelity error (NPE) for the F$_2$ molecule for the TPS methods and CC methods with respect to FCI in $mE_h$  along the PES (r from 1.2 $\AA$ to 3.1 $\AA$).}
\label{tbl:npe}
\begin{tabularx}{0.6\linewidth} {
                                  >{\centering\arraybackslash  \hsize=.6\hsize}X
                                | >{\centering\arraybackslash  \hsize=.4\hsize}X
                                   }
\hline \hline
Method   & \multicolumn{1}{l}{NPE( $mE_h$)} \\ \hline 
TPS-PT2  & 3.29                      \\
TPS-CEPA & 2.72                      \\
TPS-ACPF & 2.95                      \\
TPS-AQCC & 3.15                      \\
TPS-CISD & 4.12                      \\
CCSD     & 22.44                     \\
CCSD(T)  & 20.54                     \\ \hline \hline
\end{tabularx}
\end{table}

For multiple bond breaking cases, we can either put all bonding and anti-bonding valence orbitals in the same cluster, 
or define each cluster such that it contains a single bonding/anti-bonding pair. 

\begin{figure}
	\begin{center}
	\includegraphics[width=0.84\linewidth]{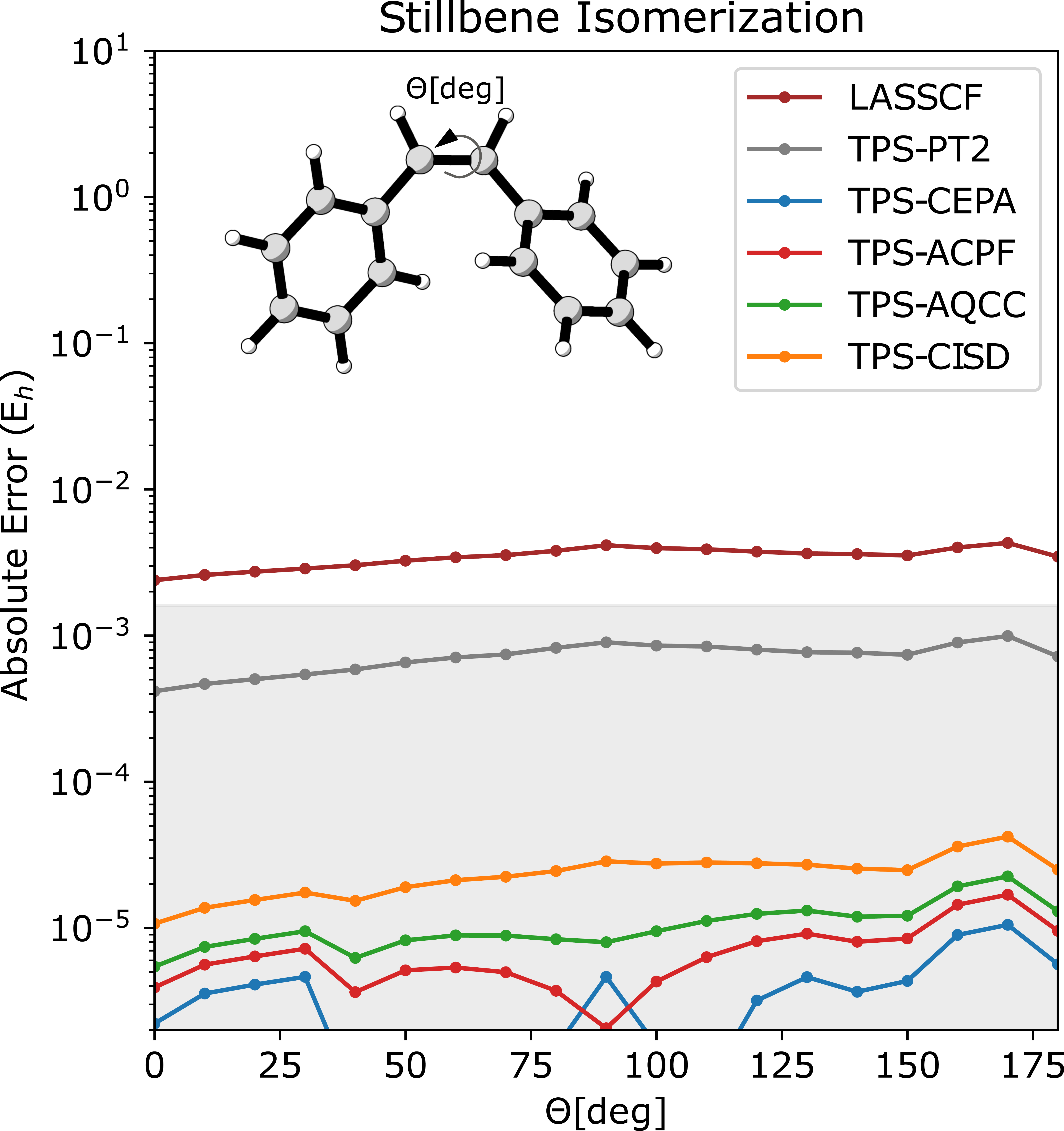}
	\caption{Absolute error with respect to the exact diagonalization inside the active space with the TPS approaches for the stillbene isomerization PES.}
	\label{fig:sb}
	\end{center}
\end{figure}

\subsection{Stillbene Isomerization}
Isomerization of stillbene\cite{Waldeck1991,Gagliardi2002,Quenneville2003,Dobryakov2012,Minezawa2011,Chaudhuri2013,Tomasello2013,Weir2020} is a challenging problem for a partition based approach such as the one presented in this work.
The active space has 10 electrons in 10 orbitals and upon localization it can be seen that each benzene unit has 4 orbitals
    each and the central double bond has 2 orbitals.
This is similar to the partitioning used by Pandharkar and coworkers for their pair-density functional theory (PDFT) based localized active space (LAS) wavefunction.\cite{Pandharkar2021}

We generate the full potential energy surface of the isomerization of \textit{cis} to \textit{trans}-stillbene by fixing the dihedral angle of the central double bond at each point along the PES and optimizing the rest of the molecule at the B3LYP/6-31G(d,p) level.
Since the active space is very small, we compare our approach to the exact diagonalization.
Because our current cMF implementation works in the MO basis, we use the variational LASSCF code, which works in the atomic orbital basis, allowing us to use an optimized core in our reference for stillbene.\cite{Hermes2020} 
The LASSCF method is similar to cMF but with an emphasis on using localized subspaces, and produces numerically identical results to cMF.

We compute the TPS-based coupled electron quantities using this system and present error with respect to the exact diagonalization in Figure \ref{fig:sb}.
All approaches, even those including the PT2 correction, give accurate results for the whole PES curve.
It can be seen that the TPS-CEPA performs best among all the options.
Even though stillbene is a small system, we can see that a mean field in the TPS basis is still not sufficient for accurate results and inter-cluster correlations need to be included.
Capturing the dynamic correlation outside the active space is also of interest and this is one of the main challenges for future studies.
One of the future direction would be to use the adiabatic connection based methods that have been recently used for capturing the dynamic correlation outside the active space using only up to two-body reduced density matrices.\cite{beran2021density,drwal2022efficient}

\begin{figure}
	\begin{center}
	\includegraphics[width=\linewidth]{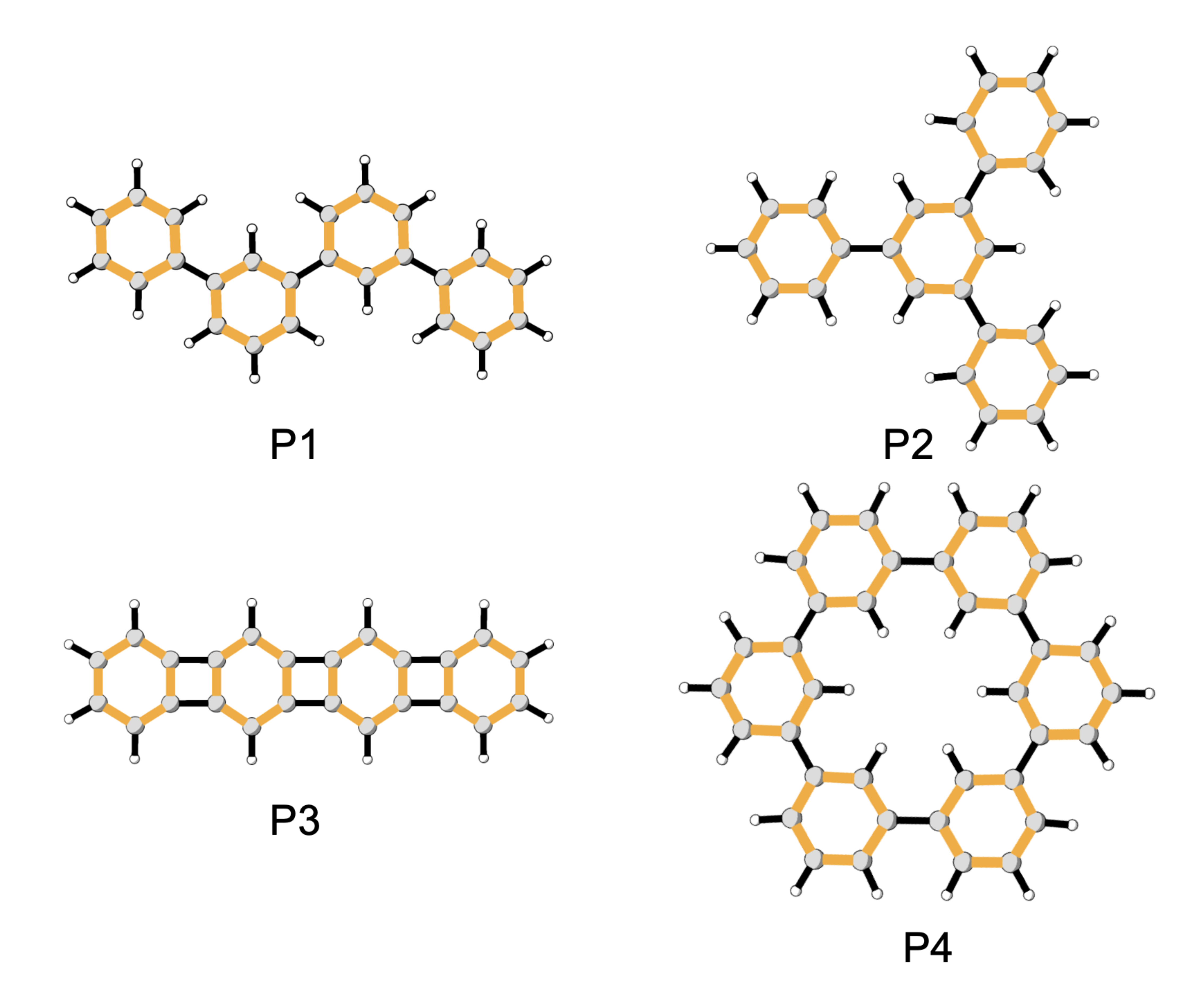}
	\caption{PAH systems considered in the study. Each benzene $\pi$ unit (highlighted in yellow) is partitioned into a cluster.}
	\label{fig:pah}
	\end{center}
\end{figure}

\subsection{Planar $\pi$-conjugated systems}

\subsubsection{PAH systems}

In this section, we investigate a few PAH systems using the TPS based coupled pair methods.
We select a few clusterable PAH systems as shown in Figure \ref{fig:pah}.
We partition each of these PAHs such that each benzene unit is a cluster with 6 orbitals. 
In the case of \textbf{P1} and \textbf{P2}, the clusters are only connected by a single C-C connection and can be considered clusterable.
\textbf{P3} and \textbf{P4} are relatively difficult for a clustered approach since \textbf{P3} has larger inter-cluster interactions and \textbf{P4} is a larger system.
We correlate only the $\pi$ orbitals and see how the method works for each of these systems.
Systems \textbf{P1}, \textbf{P2} and \textbf{P3} have an active space of (24o,24e) and \textbf{P4} has an active space with (36o,36e).
Since these active spaces are large, we compare the results to extrapolated TPSCI.\cite{abrahamSelectedConfigurationInteraction2020c}

\begin{table}[]
	
\caption{Error for each TPS method with respect to the extrapolated TPSCI (near FCI) values in  $mE_h$. The extrapolated TPSCI values are presented in the final row in Hartrees.}
\label{tbl:pah}
\begin{tabular}{l|r|r|r|r}
\hline
	     \hline
             & \multicolumn{1}{l|}{\textbf{P1}} & \multicolumn{1}{l|}{\textbf{P2}} & \multicolumn{1}{l|}{\textbf{P3}} & \multicolumn{1}{l}{\textbf{P4}} \\ \hline
	     cMF          & 15.38                   & 15.45                   & 31.41                   & 30.99                  \\
	     TPS-MP2      & 3.12                    & 3.19                    & 8.05                    & 6.57                   \\
	     TPS-CEPA     & -0.13                   & -0.07                   & -0.51                   & 0.02                   \\
	     TPS-ACPF     & -0.09                   & -0.03                   & -0.31                   & 0.12                   \\
	     TPS-AQCC     & -0.05                   & 0.01                    & -0.13                   & 0.22                   \\
	     TPS-CISD     & 0.30                    & 0.36                    & 1.53                    & 1.63                   \\ \hline
	     Extrap TPSCI & -919.6320               & -919.6314               & -915.9380               & -1377.7107             \\ \hline
	     \hline
\end{tabular}
\end{table}

We present data for all the PAH systems in Table \ref{tbl:pah}.
The difference between cMF and the extrapolated TPSCI is the inter-cluster correlation energy that needs to be accounted for.
It can be seen that in \textbf{P3} and \textbf{P4} the inter-cluster correlation energy almost doubles from that of \textbf{P1} and \textbf{P2}. 
For the TPS-MP2, as well as the TPS-CISD, the errors are even larger as the system gets more challenging in the cases of \textbf{P3} and \textbf{P4}.
However, if we instead compute the electron correlation per inter-cluster electron pair, we find that \textbf{P1}
, \textbf{P2}, and \textbf{P4} all have nearly identical correlation energies, with \textbf{P3} having nearly twice the correlation per inter-cluster electron pair arising from the increased connectivities between the clusters.
Compared to the TPS-MP2 and TPS-CISD, all the coupled pair approaches using the TPS framework give excellent results with errors consistently below 1 $mH$.
Among the three coupled pair approaches, the TPS-AQCC provides slightly better results.

\begin{figure}
\begin{center}
	\includegraphics[width=\linewidth]{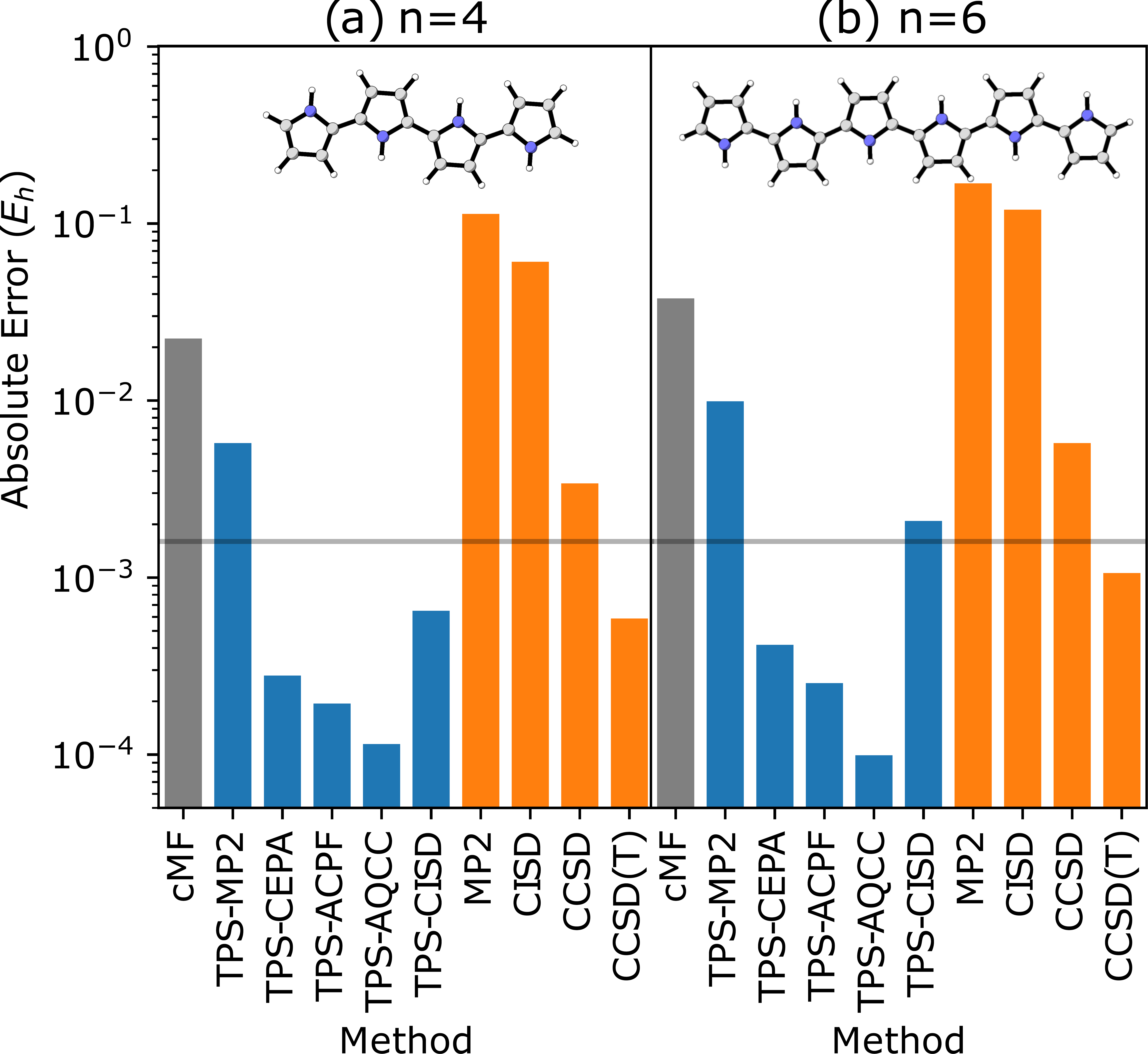}
	\caption{Comparison of the Slater determinant based method with the tensor product based approach for the polypyrrole polymer system with (a) n = 4 and (b) n = 6 pyrrole units.}
	\label{fig:ppy_data_lcc}
\end{center}
\end{figure}

\subsubsection{Polypyrrole}
Polypyrrole (\textbf{PPy}) polymer units have potential applications in biomaterials and as molecular wires. \cite{Smela2003,Pernaut2000,huang_nanostructured_2016,yuan_photocatalytic_2019} 
Unlike the PAH systems, the \textbf{PPy} system has a heteroatom (N). 
For the TPS method, we partition this system such that each pyrrole unit is a cluster.
Here we present the \textbf{PPy} molecule in its reduced neutral form with 4 and 6 pyrrole units as an example.
We compute the correlation energy of the active space defined by the $\pi$ space in a similar manner to the PAH systems.
Partitioned in this way, the active space for a single pyrrole unit is (5o,6e), leading to a (20o,24e) and (30o,36e) active space for the 4 and 6 unit \textbf{PPy} systems.
We use the extrapolated TPSCI results as the reference value.

We present the data for both the systems in Figure \ref{fig:ppy_data_lcc}. 
As seen from the plots, traditional single reference approaches like MP2, CISD and CCSD have errors larger than chemical accuracy.
Meanwhile all of the coupled pair TPS approaches have very small errors.
Among the three coupled pair approaches, the TPS-AQCC provides the most accurate results.
It is interesting to note that the TPS mean field reference is already better than the determinant based MP2 and CISD. 

\section{Conclusion}
The CEPA family of methods often demonstrate either high accuracy when weakly correlated or dramatic failure when near-degeneracies arise. 
Despite the well known deficiencies, CEPA remains important as a potential tool due to two properties: (1) it is a size-extensive framework, and (2) it can be extended to multi-reference cases easily.
In this work, we generalize the CEPA framework within a  tensor product state basis.
This provides a technique for using CEPA methods to recover the inter-cluster correlations from a cluster mean-field reference state. 
Using the TPS basis, the CEPA equations are much more stable and gives accurate results for the PES scan for single bond breaking.
We further apply this formalism to stillbene isomerization and a variety of $\pi$-conjugated systems where very accurate results are obtained.

The present study is preliminary, with many important future directions available for making the TPS-CEPA method applicable to general systems. 
Firstly, the data for the present study has been generated by modifying a sparse TPS code.
We use a threshold in the first order interacting space of the Hamiltonian to truncate the wavefunction.
An improved implementation with a dense framework can help us target much larger systems, by leveraging efficient BLAS routines rather than element-wise sparse operations.
Another important future direction is to capture dynamic correlation from large virtual space orbitals.
It would also be interesting to extend this approach to multi-reference TPS framework.


\section{Supporting Information}
The effect of clustering in the F$_2$ molecule is discussed.
We have also included the orbital clustering for CH$_4$,F$_2$ and stillbene.

\section{Acknowledgements}
This research was supported by the  National Science Foundation (Award No. 1752612). 
The authors acknowledge Advanced Research Computing at Virginia Tech for providing computational resources and technical support that have contributed to the results reported within this paper. 

\providecommand{\latin}[1]{#1}
\makeatletter
\providecommand{\doi}
  {\begingroup\let\do\@makeother\dospecials
  \catcode`\{=1 \catcode`\}=2 \doi@aux}
\providecommand{\doi@aux}[1]{\endgroup\texttt{#1}}
\makeatother
\providecommand*\mcitethebibliography{\thebibliography}
\csname @ifundefined\endcsname{endmcitethebibliography}
  {\let\endmcitethebibliography\endthebibliography}{}

\end{document}


\title{Supporting Information: Coupled electron pair-type approximations for tensor product state wavefunctions} 

\author{Vibin Abraham}
\email{avibin@umich.edu}
\affiliation{Department of Chemistry, University of Michigan, Ann Arbor, MI 48109, USA}
\author{Nicholas J. Mayhall}
\email{nmayhall@vt.edu}
\affiliation{Department of Chemistry, Virginia Tech,
Blacksburg, VA 24060, USA}

\maketitle

\beginsupplement
\section{Effect of Cluster size for F$_2$}
\subsubsection{Effect of clustering}
For the TPS based approach, we partition the system such that the bonding and anti-bonding pairs are in the same cluster leading to 8 clusters. 
Another clustering option would be to include the dynamic correlation part for the valence orbitals by groupling the orbitals of same angular momentum but different principle quantum numbers as one cluster.
We will end up with 4 orbitals per cluster in 4 different clusters. 
These clusters can be used for any diatomic and we have previously shown how these clustering choices are good for the N$_2$ molecule.\cite{Abraham2020}

\begin{itemize}
	\small
	\item 8c:(2s), (3s), (2p$_z$), (3p$_z$), (2p$_x$), (3p$_x$), (2p$_y$), (3p$_y$) 
	\item 4c:(2s, 3s), (2p$_z$, 3p$_z$), (2p$_x$, 3p$_x$), (2p$_y$, 3p$_y$) 
\end{itemize}

\begin{figure}
	\begin{center}
	\includegraphics[width=0.7\linewidth]{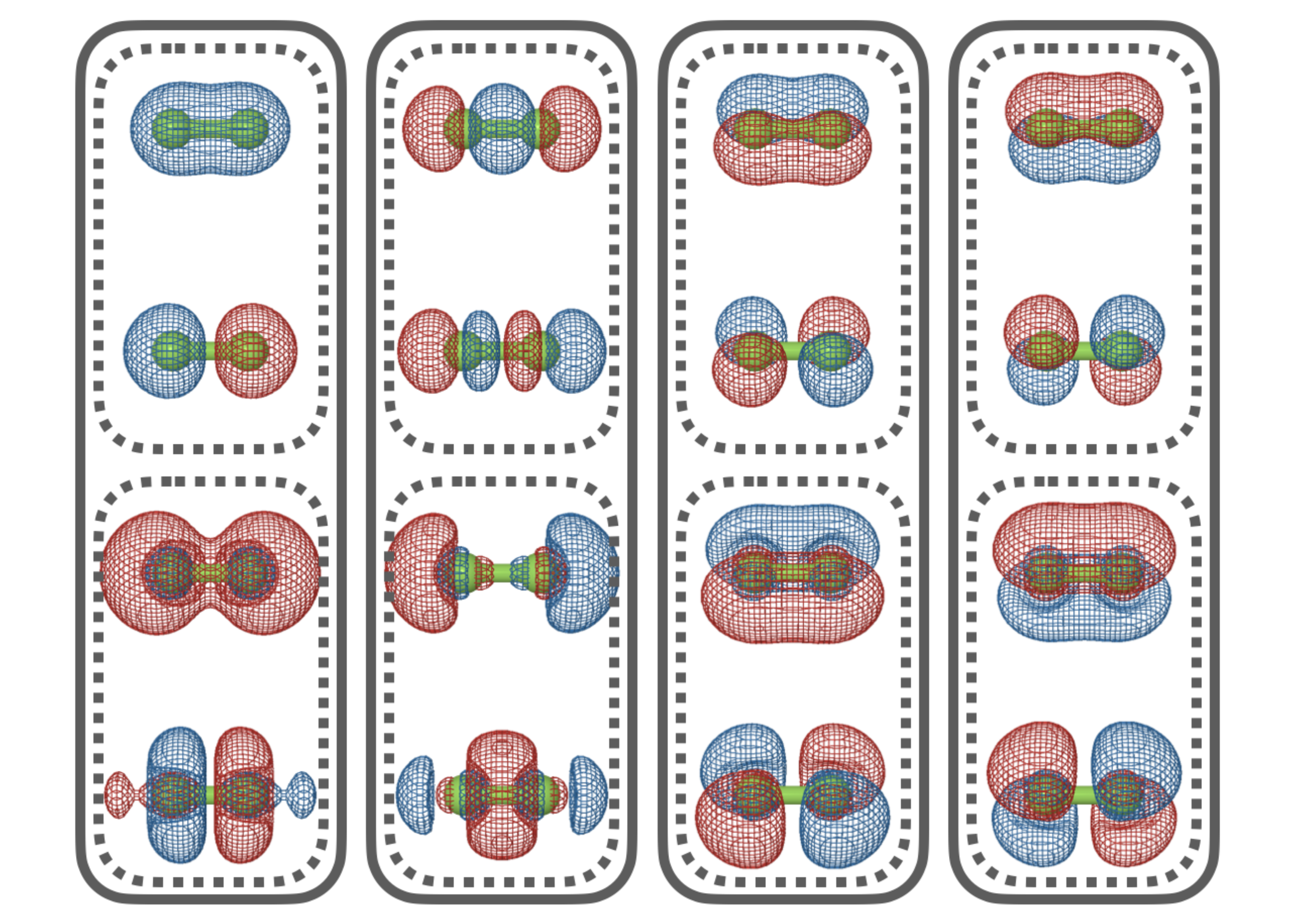}
	\caption{The two clustering considered for the F$_2$ molecule.
	The dotted line corresponds to 8c clustering and the solid line corresponds to 4c clustering.
	The data presented in the main text are generated using the 4c clustering.}
	\label{fig:f2_mo}
	\end{center}
\end{figure}

\begin{figure}
	\begin{center}
	\includegraphics[width=0.7\linewidth]{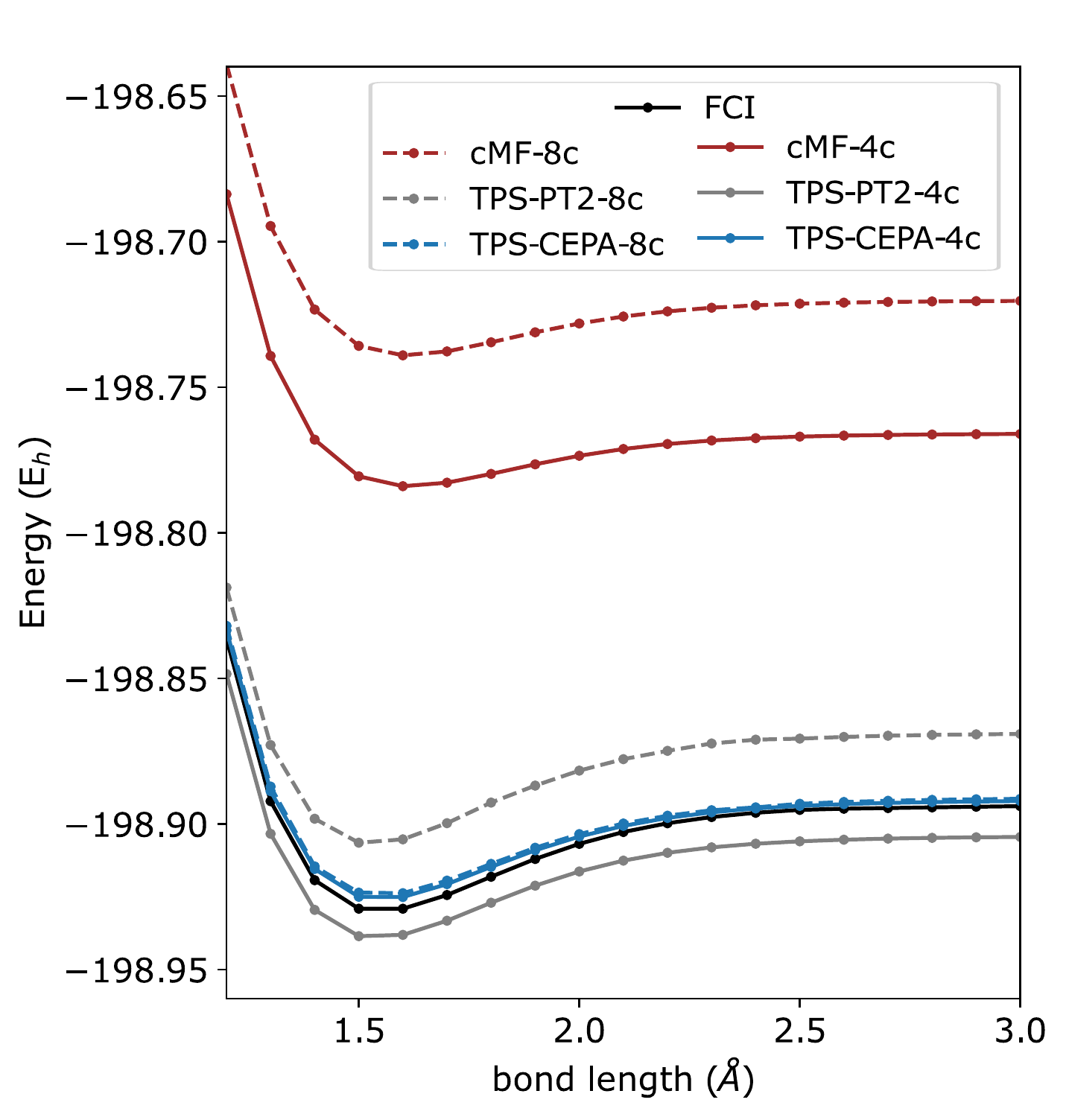}
	\caption{Comparison of the two clustering schemes 8c and 4c along the PES curve of F$_2$. The dotted(solid) lines correspond to 8c (4c) data.}
	\label{fig:f2_clu}
	\end{center}
\end{figure}

We present data for the two clustering discussed above in Figure \ref{fig:f2_clu}.
As can be seen the cMF values are much higher for both 8c and 4c than the actual PES curve, suggesting that the 
    correlation in between clusters are important for this system.
The cMF for the 4c is relatively better since some of the intercluster correlation effects in 8c is already included in 4c.
For the PT2 correction, we see that the 4c and 8c curve are qualitatively accurate, but the 4c curve overestimates the FCI energy
Therefore we need a more accurate estimate for the inter-cluster correlation than a simple PT2 correction.
Finally, the TPS-CEPA for both 8c as well as 4c look almost identical with 4c slightly better.
TPS-CEPA corrects for the inter-cluster correlation efficiently for both 8c as well as 4c. 
Hence TPS-CEPA can be used as an efficient approximation for capturing the dynamic correlation energy for these systems.

\section{Clustering choice for CH$_4$ and Stillbene}

\begin{figure}
	\begin{center}
	\includegraphics[width=0.7\linewidth]{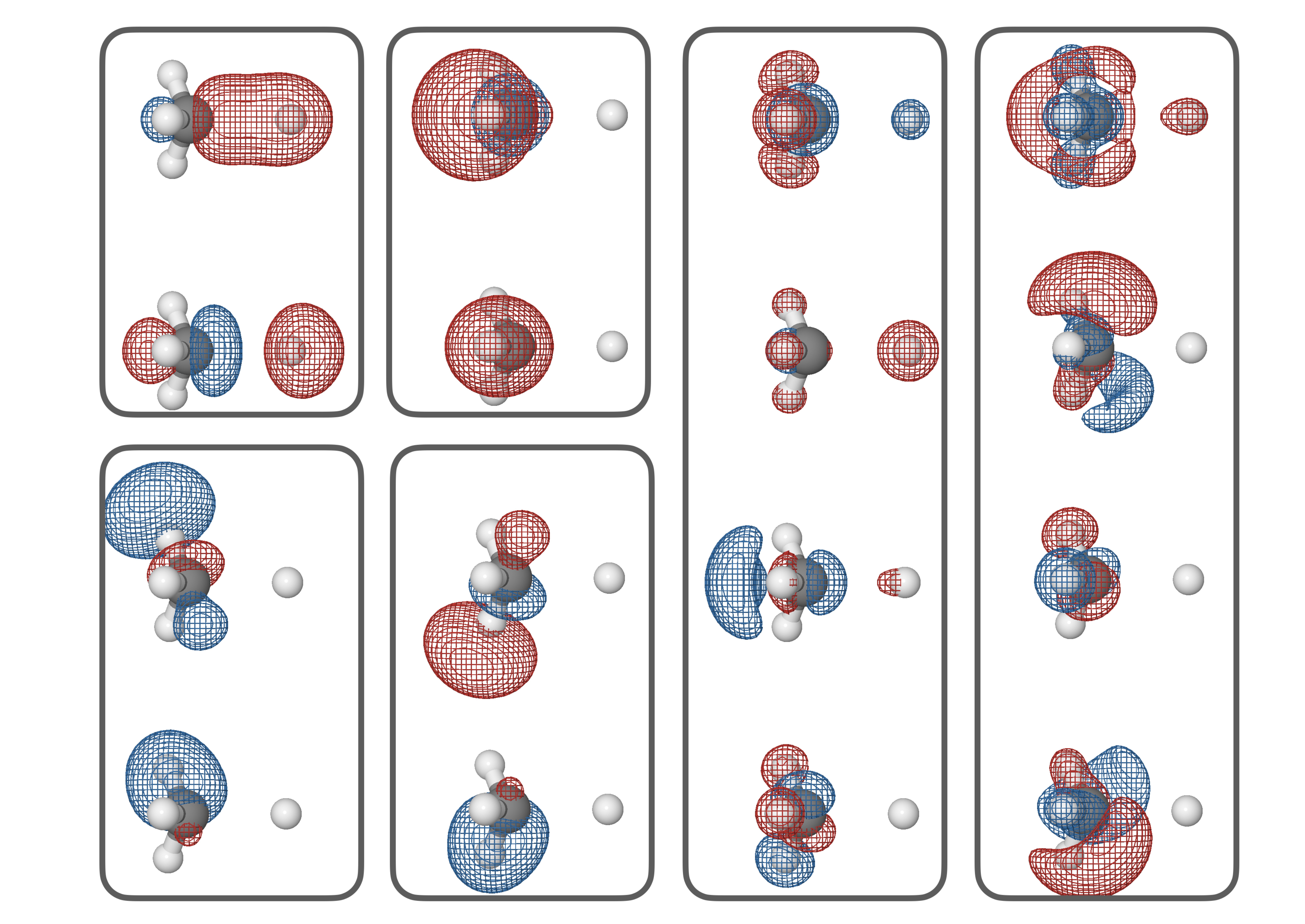}
	\caption{The molecular orbital clustering for the CH$_4$ single bond breaking.}
	\label{fig:ch4}
	\end{center}
\end{figure}

\begin{figure}
	\begin{center}
	\includegraphics[width=0.7\linewidth]{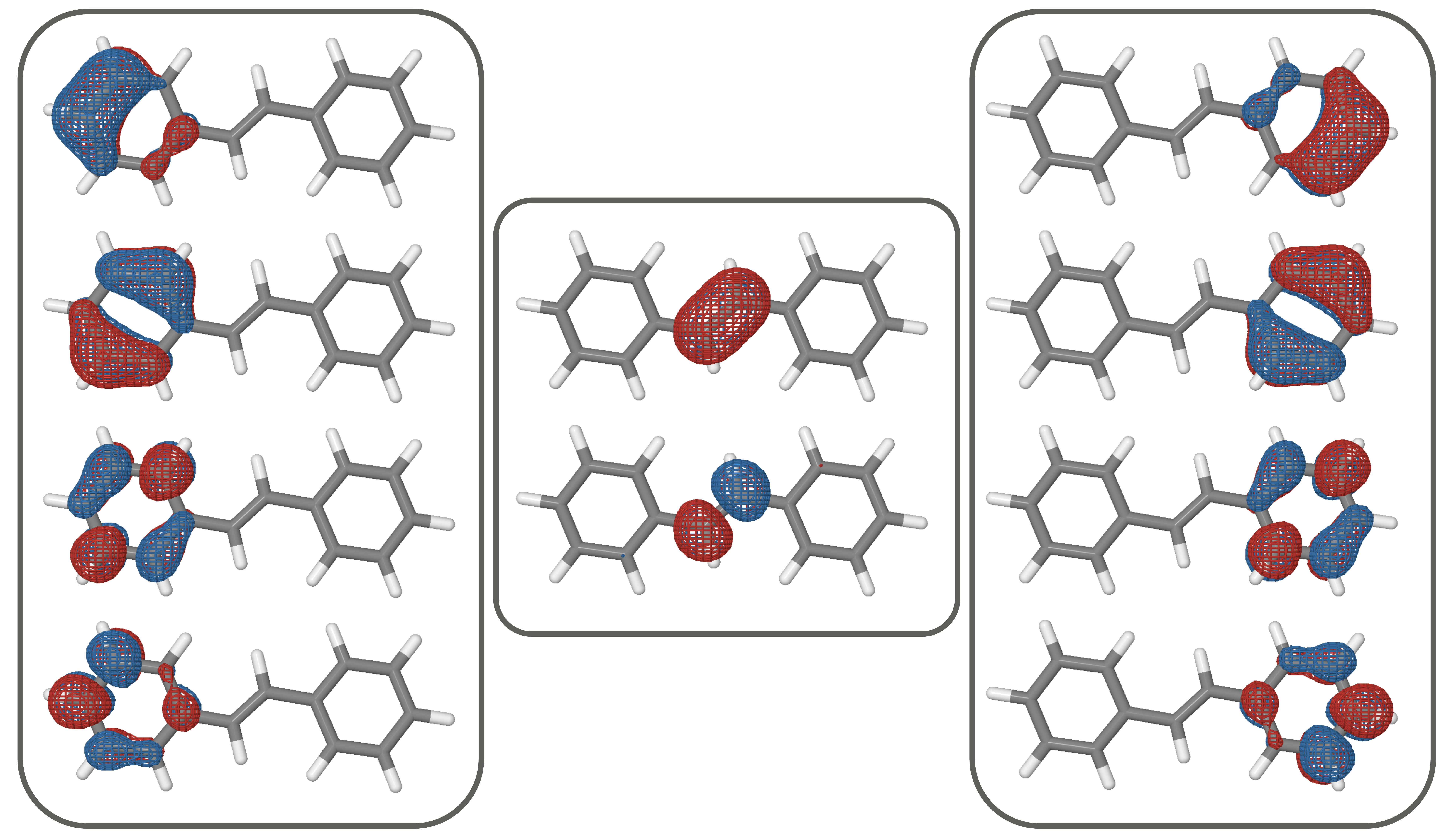}
	\caption{The localized molecular orbitals of the \textit{trans}-stillbene. The orbitals have been partitioned into three clusters for the present study.}
	\label{fig:sb_mo}
	\end{center}
\end{figure}